
\magnification=\magstep1
\parskip 0pt
\def\({[}
\def\){]}

\def\rjustline#1{\line{\hss#1}}

\overfullrule 0pt
\rjustline{WIS-93/12/Jan-PH}
\bigskip
\centerline{Physics at B-Factories (A Theoretical Talk)}

\centerline{Harry.~J.~Lipkin}
\vskip 12pt
\centerline{\it Department of Nuclear Physics}
\centerline{\it Weizmann Institute of Science}
\centerline{\it Rehovot\ \ 76100\ \ \ ISRAEL}
\vskip 12pt
\centerline{and}
\vskip 12pt
\centerline{\it School of Physics and Astronomy}
\centerline{\it Raymond and Beverly Sackler Faculty of Exact Sciences}
\centerline{\it Tel Aviv University, Tel Aviv, Israel}

\centerline{January 2, 1993}
\vskip0.3in
\baselineskip 12pt

\medskip

      The main motivation for B factories is the investigation of CP violation.
We consider two questions: 1) Why has there been essentially no experimental
progress in CP violation in the 28 years since its discovery in 1964? 2) Why
do we expect better results from B factories?

\noindent{\twelvepoint 1.\  {\caps Introduction - The Difference Between
P and CP}}

\noindent
One of the greatest missed opportunities in my scientific career
occurred in the summer of 1964 when Jim Smith came up from Urbana to
Argonne where I was spending the summer and told me that they had seen
events in their Brookhaven experiment suggesting that the long-lived
neutral kaon decayed into two pions. He asked me whether there was any
theoretical explanation and in particular whether this result could be
explained without assuming CP violation.

\noindent
I could have immediately started writing a great paper on CP
phenomenology discussing the implications of $K_L \rightarrow 2\pi$ and
including the $\phi$ factory proposal\REF{\HJLPHI}{Harry J. Lipkin,
Phys. Rev. 176 1715 (1968).}$\({\HJLPHI}\)$
to measure what is now called $\epsilon'$.
I had been using kaon physics and EPR-correlated decays of kaon pairs as
excellent exercises in teaching quantum mechanics. Instead
I told Jim that there was no sensible theoretical explanation for
this result and that they should probably check out their experiment
further. Their paper appeared\REF{\Illinois}{A. Abashian et al.,
Phys. Rev. Lett. 13, 243 (1964)}$\({\Illinois}\)$
two weeks after the Cronin-Fitch paper\REF{\Cronin}{J.H. Christensen et
al., Phys. Rev. Lett. 13, 138 (1964).}$\({\Cronin}\)$
and I published my $\phi$ factory paper four years later$\({\HJLPHI}\)$.

\noindent
In contrast to parity violation, where the
discovery\REF{\Ambler}{E. Ambler et al., Phys. Rev. 105, 1413
(1957).}$\({\Ambler}\)$
was immediately followed by
many experiments and the new theoretical
description was soon clear, the only new experimental evidence for CP
violation in the past 28 years has been in
\noindent
refinements of the original experiments studying neutral kaon decays
and getting better values for the crucial basic parameters called
$\epsilon$  and $\epsilon'$. But the data are still not good enough to
distinguish between two classes of theoretical models predicting that
$\epsilon'$ is finite or zero.

\noindent
In 1956 we did a $\beta$-ray polarization
experiment\REF{\ADSLIP}{A. de-Shalit et al., Phys. Rev. 107, 1459
(1957).}$\({\ADSLIP}\)$
confirming that parity was violated and showing that
the weak current coupled dominantly to left-handed electrons.
This experiment was so simple that even Lipkin could do it
under the primitive conditions at the Weizmann Institute at that time.
Why was it not done earlier?
Anyone who started this experiment at the same time that C. S. Wu et al
started their parity experiment$\({\Ambler}\)$
would have obtained his results first and discovered parity violation.
But nobody bothered. The community was brainwashed
by theorists insisting that parity violation predicted by two
young crazy theorists was complete nonsense. Nobody would do a simple
experiment to find a 100\% parity violation where a negative result
would prove nothing. Ambler et al$\({\Ambler}\)$ did a difficult
experiment sensitive enough to
detect parity violations of a few per cent where a negative result could
shoot down this crazy theory. But once the effect was found, there
was immediate tremendous progress in both experiment and theory.

\noindent
The discovery of CP violation in kaon decay was
completely different.
There was no theoretical motivation for a search, and
no real progress for
\noindent
28 years after the discovery.
We now examine the underlying reasons for this difference.

\noindent
Two types of experiments can detect CP violation:
(1) Charge asymmetry; e.g.
$\Gamma(B^+ \rightarrow X) \not=  \Gamma(B^- \rightarrow \bar X)$;
(2) Neutral meson mixing, where mass eigenstates like $K_L$ and $K_S$
are shown not to be $CP$ eigenstates.
Charge asymmetries have not yet been found. We shall see below why they
are hard to find and why there is hope of finding them in $B$ decays.
In the kaon system
the lifetimes of the two mass eigenstates are so different that a
pure $K_L$ beam can be created simply by waiting. Its $CP$ properties are
determined by observing its decay into $CP$ eigenstates. This is the only
way that CP violation has been observed to date. It is not possible in
$B$ and $D$ systems where the lifetimes of the two mass eigenstates are
essentially equal.

\noindent

\medskip
\noindent{\twelvepoint 2.\ {\caps Why CPT Makes Charge Asymmetries Hard
To Find}}

\medskip

\noindent
We assume CPT invariance. Then the
total decay widths of charge conjugate heavy meson states; e.g. $B^+$ and
$B^-$ are required to be equal.
If CP is violated
partial widths for decays into charge conjugate
exclusive modes can be different.
But CPT requires all these different partial widths to add up to the same
total widths for two charge conjugate mesons. We now show how this
apparent miracle follows from simple general dynamical arguments.
\smallskip
\noindent{\twelvepoint 2.1 \ How CPT Makes Total Widths Equal}

\noindent
We first note how CPT invariance requires equal total widths for
$B^{\pm}$ states.
In s-wave $K\pi$ elastic scattering in the energy region near the $B$
mass the $B$ appears as a very narrow resonance
with a Breit-Wigner shape. Since
the elastic s-wave $K^+\pi^o$ cross section goes into the elastic
s-wave  $K^-\pi^o$ cross section under CPT at all energies, the shapes
of the two resonances at the $B$ mass must be equal. Thus
$$
\vbox{\eqalignno{
\sigma_{el,s}(K^+\pi^o) & = \sigma_{el,s}(K^-\pi^o)
&(QQ2.1a)  \cr
\Gamma_{tot}(B^+) & = \Gamma_{tot}(B^-)
& (QQ2.1b)  \cr}}   $$

\smallskip
\noindent{\twelvepoint 2.2 \ Golden Rule Makes Partial Widths Equal}

\noindent
We next note that CPT invariance and the hermiticity of the CP-violating
weak interaction $H_{wk}$ require the partial widths for
$B^{\pm}$ decays
into a pair of charge conjugate exclusive final states also to be equal
in first-order perturbation theory, where the decay rate is described by
the Fermi Golden Rule. For any such pair
denoted by $f^\pm$, the golden rule gives
$$ W_{B^-\rightarrow f^\pm}\approx {{2\pi}\over{\hbar}}
|\bra{f^\pm}H_{wk}\ket{B^\pm}|^2
 \rho(E_f)  \eqno(QQ2.2b)    $$
where $\rho(E_f)$ is the density of final states.
But from CPT and hermiticity,
$$ {{|\bra{f^-}H_{wk}\ket{B^-}|}\over
{|\bra{f^+}H_{wk}\ket{B^+}|}}= {{CPT |\bra{B^+}H_{wk}\ket{f^+}|}
\over{|\bra{B^+}H_{wk}\ket{f^+}^*|}} = 1
 \eqno(QQ2.3a)    $$
$$ W_{B^+\rightarrow f^+}\approx  W_{B^-\rightarrow f^-}
 \eqno(QQ2.3b)    $$
Thus a charge asymmetry in partial widths can only occur in channels
where the golden rule breaks down; i.e. where corrections beyond first
order are important.
Weak radiative corrections are negligible; thus only strong interaction
rescattering corrections can produce a significant violation of the
golden rule and a significant CP violation effect in exclusive partial
widths. However, the asymmetric partial widths must
still add up to give equal total widths.

\smallskip
\noindent{\twelvepoint 2.3 \  How Charge Asymmetries can Occur}

\noindent
A simple example of golden rule breakdown is seen in the
effect of strong charge exchange scattering
on the relations (QQ2.2)
$$ {{W_{B^+\rightarrow K^+\pi^o}}\over{
W_{B^-\rightarrow K^-\pi^o}}} = {{
|S_{el}M(K^+\pi^o) + S_{cex}M(K^o\pi^+)|^2}\over{
|S_{el}M(K^-\pi^o) + S_{cex}M(\bar K^o\pi^-)
|^2}}
 \eqno(QQ2.4a)    $$
where
$$ M(f^\pm) \equiv \bra{f^\pm}H_{wk}\ket{B^\pm}
 \eqno(QQ2.4b)    $$
and $S_{el}$ and $S_{cex}$ depend respectively on the $CP$-conserving
strong interaction elastic and charge exchange scattering amplitudes and
are the same for the two charge-conjugate transitions.
The weak matrix elements can have different relative phases if there is s
$CP$ violation.

\noindent
Thus a CP-violating asymmetry can be observed when
final state rescattering couples
two final states whose weak matrix elements have different weak phases.
The total decay widths involve transitions to all other final states
including those like $K^o\pi^+$ and $\bar K^o\pi^-$ which involve the
same weak matrix elements and charge exchange strong rescattering as in
the transitions (QQ2.4).
This rescattering back and forth between states is constrained
by the unitarity of the strong-interaction S matrix to preserve the
equality of the total widths of the $B^+$ and $B^-$ decays.
This is simply illustrated in a toy model where all
$B^\pm$ decays go into the $K\pi$ modes,
and there are only two independent final states.
The two isospin eigenstates with I=1/2 and I=3/2 constitute a
complete set for expanding the final states.
Since the strong interactions conserve isospin,
there is no final state rescattering between states of different isospin
and the Golden Rule and eqs. (1.4) hold for these states. Thus
 $$|A\{(B^+ \rightarrow (K  \pi)_I\}| =
|A\{B^- \rightarrow  (\bar K  \pi)_I\}| \eqno(QQ2.5)$$
where $(K  \pi)_I$ denotes the $K \pi$ isospin
eigenstate with isospin I/2.
We now consider the decay amplitudes for the
$K^+\pi^o$, $K^o\pi^+$, $K^-\pi^o$ and  $\bar K^o\pi^-$ decay modes,
The decay amplitude into any final state $\ket{f^\pm}$
can be written in terms of the isospin amplitudes (QQ2.5),
a CP-violating weak phase denoted by $W_I$
which is $opposite$ for charge-conjugate $B^\pm$ and a strong
CP-conserving phase denoted by $S_I$ which is the $same$
for charge-conjugate amplitudes.
 $$A\{B^\pm \rightarrow f^\pm \} =
$$ $$ =
 \sum_{I=1,3} C^{f}_I|A\{B^\pm \rightarrow (K  \pi)_I\}|
\cdot e^{\pm iW_I}e^{iS_I}
\eqno(QQ2.6)  $$
where $ C^{f}_I$  denotes isospin Clebsch-Gordan
coefficients.
The charge-conjugate amplitudes are now seen to be not necessarily
equal. The CP asymmetry for the individual charge states is given by
$$|A\{B^+ \rightarrow K^+\pi^o \}|^2 - |A\{B^- \rightarrow K^-\pi^o \}|^2
= $$
$$ = -4 C^{f}_1 C^{f}_3|A_1A_3|
\sin(W_1-W_3)\sin(S_1-S_3)
\eqno(QQ2.7a)  $$
$$|A\{B^+ \rightarrow K^o\pi^+ \}|^2 - |A\{B^- \rightarrow
\bar K^o \pi^- \}|^2 = $$
$$ = 4 C^{f}_1 C^{f}_3|A_1A_3|
\sin(W_1-W_3)\sin(S_1-S_3)
\eqno(QQ2.7b)  $$
where the Clebsch-Gordan coefficients $C^{f}_I$ are defined for the
final states $K^\pm \pi^o$ and the Clebsch-Gordan orthogonality relation
requires $C^{f}_1 C^{f}_3$ to be equal and opposite for the two states
(QQ2.7a) and (QQ2.7b).
The asymmetries vanish unless $W_1 \not= W_3$ $and$ $S_1
\not= S_3$, have opposite signs for the two charge states.
and cancel in the total decay rates as expected from CPT.

\noindent
The $CP$ asymmetry of unequal partial widths for charge
conjugate exclusive decay modes results from interference terms between
amplitudes coming from different weak interaction diagrams which can have
different relative phases. These phases depend upon the interplay between the
weak interaction phases and the strong phases from final state interactions.
In the general case where many exclusive channels are all coupled by the final
state strong rescattering, explicit calculations are difficult. However, all
the physics needed is in one general feature of FSI displayed in this model
with two final states; namely that the FSI are
described exactly by a unitary $S$ matrix which contains all the strong
interaction dynamics.

\noindent
A more intuitive picture is given by eq. (QQ2.4) which refers to elastic
and charge exchange scattering rather than to isospin eigenstates. When
$S_3 = S_1$ the strong charge exchange scattering amplitude vanishes.
Instead of expressing the physics in terms of strong phases, which are
not intuitively evident, one can note that unequal partial widths for
charge-conjugate exclusive decay modes can result from
strong interaction rescattering between different intermediate states
produced by different weak interaction diagrams with different weak
phases.

\medskip
\noindent{\twelvepoint 2.4 \  Summary of CPT Constraints}

\noindent
We can now summarize the situation as follows:

\noindent
1. CPT invariance requires equal total decay widths of charge
conjugate states.

\noindent
2. In the Fermi-Golden-Rule approximation
of first-order time-dependent perturbation theory, CPT invariance
and the hermiticity of the interaction
require equal partial decay widths into charge conjugate exclusive
channels.

\noindent
3. CPT and hermiticity thus allow CP-violating charge-asymmetric decays
only in cases where the Fermi golden rule does not apply.

\noindent
4. In the approximation where the CP-violating weak interaction is
treated only in first order, and the strong interaction which is treated
to all orders conserves CP,
the decays into final states which are eigenstates of the strong
interaction $S$ matrix are described by the Fermi Golden Rule.
Thus CPT invariance and the hermiticity of the interaction
require that the partial decay widths into charge conjugate exclusive
channels be equal.

\noindent
5. A necessary condition for CP-violating
charge-asymmetric decay rates is that the decay amplitudes must have
contributions from at least two strong interaction eigenstates for which
both the strong phases and the weak phases are different. This
condition is also sufficient except in special cases of accidental
cancellations which can arise when there are many contributing strong
interaction eigenstates.

\noindent
6. The simple physics underlying this condition is that strong
interactions couple different final states, thereby allowing two
different weak transitions to contribute to the same final state
via two different intermediate states;
e.g. to $K^+ \pi^o$ via the intermediate states $K^+ \pi^o$
and $K^o \pi^+$ and strong elastic and charge-exchange scattering.
These contributions are coherent and
therefore depend upon the relative phase of the two weak transitions
which can be CP-violating. However, since the strong interaction S matrix
is unitary, the ``off-diagonal"contribution to the
$K^+ \pi^o$ final state via the intermediate state
$K^o \pi^+$ must be cancelled in the total decay width by
the ``off-diagonal"contribution to the
$K^o \pi^+$ final state via other intermediate states like
$K^+ \pi^o$.

\noindent
We can now ask how these considerations can help look for promising CP
asymmetries both within and beyond the standard model. We need two
eigenstates of the strong interaction S matrix and contributions from
two different weak interaction diagrams which can have different phases.
In our toy model the two strong eigenstates have different isospins.
The quantum number (I=3/2) is exotic; it cannot be produced via a diagram
like a penguin which goes via an intermediate state containing only one
$q \bar q$ pair and gluons. There can be contributions
from two weak interaction diagrams. one of which, the penguin, goes via
an intermediate state which has only a single $q\bar q$ pair and
gluons, and therefore cannot have exotic flavor quantum numbers.
The tree diagram contributes to both (I=1/2) and (I=3/2) amplitudes; the
penguin only to I=1/2. Thus if the strong phases are different for
I=1/2 and 3/2 and if the tree and penguin diagrams have different weak
phases $CP$ violation can appear as a charge asymmetry. Other diagrams
coming from physics beyond the standard model and which like the penguin
turn the heavy quark into a light quark with the same electric charge
can also play the same role as the penguin. It is therefore of interest
to investigate possibilities for such neutral heavy to light
transitions, both theoretically and experimentally.

\medskip
\noindent{\twelvepoint 3.\ {\caps CP Violation in mixed neutral mesons}}

\noindent
If CP is conserved, the decay into a CP eigenstate defines a basis of
linear combinations of $B^o$ and $\bar B^o$ which are both mass and
CP eigenstates.
These states will be stationary and not undergo mixing as a function of
time. If CP is violated, the decay into a CP eigenstate defines a basis
of states which are not mass eigenstates and can oscillate as a function
of time. Detecting the CP violation in the neutral $B$ system involves
determining whether or not the states defined by decay into a CP
eigenstate
are mass eigenstates, or whether both of the mass eigenstates
decay into the same CP eigenstate as in
$K_L \rightarrow \pi \pi $ and
$K_S \rightarrow \pi \pi $.

\noindent
The differences between the lifetimes of the two B-meson and the two
D-meson eigenstates are very small.
In the kaon system
the lifetime of the $K_S$ is determined by the dominant
$2\pi$ decay mode.
The lifetime of the $K_L$ is much longer
because no other decay mode has a comparable partial width.
The $B^o$ - $\bar B^o$ and $D^o$ - $\bar D^o$
systems have no such dominant decay mode. The principal decay
modes come in pairs with equal partial widths, one coupled to
$B^o$ and one to $\bar B^o$ and similarly for
$D^o$ - $\bar D^o$.
The lifetimes of the two eigenstates of the oscillating system
are very nearly equal.

\smallskip
\noindent{\twelvepoint 3.1\ A Pauli Spin Matrix Description of
the Oscillating $B^o$ - $\bar B^o$ System}

\noindent
It is convenient to define a ``quasispin" algebra, analogous to isospin.
The $B^o$ and $\bar B^o$ are classified in a doublet and
called ``spin up" and ``spin down", while the operators acting in
this two-dimensional space are described as linear combinations of
Pauli spin matrices\REF{\TDLee}{T. D. Lee and C. S. Wu,
Ann. Rev. Nuc. Sci. 16, 511 (1966)}\REF{\Bspin}{Harry J. Lipkin,
Argonne preprint ANL-HEP-PR-88-66
Submitted to Nuclear Physics B (1988) }
\REF{\PEPRspin}{Harry J. Lipkin,
Physics Letters B219, 474 (1989)}$\({\TDLee,\Bspin,\PEPRspin}\)$.
We define the
``$Q$-spin" matrices denoted by
$q_x$,
$q_y$ and
$q_z$, with the same form as Pauli and isospin
matrices and
choose the $z$-axis so that the $B^o$ and $\bar B^o$ states are
eigenstates of $q_z$.
Then
$$
\vbox{\eqalignno{
 q_z \ket{B^o} = \ket{B^o};&     ~ ~ ~ ~ ~
 q_z \ket{\bar B^o} = - \ket{\bar B^o} ~ ~ ~ &(QQ3.1a) \cr
 q_x \ket{B^o} = \ket{\bar B^o};&     ~ ~ ~ ~ ~
 q_x \ket{\bar B^o} = \ket{B^o} ~ ~ ~ &(QQ3.1b) \cr
 q_y \ket{B^o} = i \ket{\bar B^o};&    ~ ~ ~ ~ ~
 q_y \ket{\bar B^o} = -i \ket{B^o} ~ ~ ~ &(QQ3.1c) \cr
    \bra {B^o} q_y \ket{B^o} =
  &  \bra {\bar B^o} q_y \ket{\bar B^o}
  = \bra {B^o} q_x \ket{B^o} =  &  \cr
  = & \bra {\bar B^o} q_x \ket{\bar B^o}  = 0  & (QQ3.1d) \cr}}
$$
Any linear combination of $B^o$ and $\bar B^o$
can be expressed as a state polarized in some direction in this
quasispin space. The most general pair of
orthogonal states in the two-dimensional $B^o \bar B^o$ space,
\noindent
$$ \ket{B_\mu} =
e^{i{{\theta}\over{2}}}\cos({{\alpha}\over{2}}) \ket{B^o} +
e^{-i{{\theta}\over{2}}} \sin({{\alpha}\over{2}}) \ket{\bar B^o}
\eqno(QQ3.2a)  $$ $$
\ket{B_\nu} = e^{i{{\theta}\over{2}}}\sin({{\alpha}\over{2}}) \ket{B^o} -
e^{-i{{\theta}\over{2}}} \cos({{\alpha}\over{2}}) \ket{\bar B^o}
\eqno(QQ3.2b)   $$
are the eigenstates of the quasispin projections on an
axis with polar co-ordinates $(\alpha, \theta) $ with respect to the
$z$ axis and $\theta = 0$ in the $+x$ direction.

\noindent
The mass eigenstates are required by CPT to be linear combinations of
$B^o$ and $\bar B^o$ with equal magnitudes. In the approximation of
equal lifetimes for the mass eigenstates, which we henceforth use in
this treatment, the mass eigenstates are orthogonal
and define a direction in quasispin space normal to the $z$ axis.
We choose this as the x direction. Thus
the mass eigenstates which we denote by $B_2$ and $B_1$ are eigenstates
of $q_x$
$$
\vbox{\eqalignno{
\ket{B_2} =& (1/\sqrt 2)( \ket{B^o} + \ket{\bar B^o})
&~ ~ ~  (QQ3.3a)  \cr
\ket{B_1} =& (1/\sqrt 2)(  \ket{B^o} - \ket{\bar B^o})
&~ ~ ~  (QQ3.3b)  \cr}} $$
$$ \vbox{\eqalignno{
q_x \ket{B_2} = \ket{B_2}; \ \ \ q_x \ket{B_1} &= - \ket{B_1}
 & (QQ3.4a) \cr
 \bra {B_2} q_y \ket{B_2} =
    \bra {B_1} q_y \ket{B_1} & =
    \bra {B_2} q_z \ket{B_2} =  \cr
    \bra {B_1} q_z \ket{B_1} & = 0
 & (QQ3.4b) \cr}} $$

\noindent
In the approximation of equal lifetimes the time evolution operator for
any oscillating state can be factorized into an exponential decay
factor which is the same for all states and a time-dependent unitary
transformation which is simply a real quasispin
rotation about the $x$ axis. In contrast to the kaon case,
the decay and the mixing are completely decoupled from one another and
the time evolution of any neutral $B$ meson state $B(0)$ at time t=0
can be written
$$ \ket {B(t)}
= e^{-{\Gamma\over 2} t}\cdot
e^{-iM_o t} \cdot e^{-i{{\Delta m}\over{2}}\cdot q_x t} \ket {B(0)}
\eqno(QQ3.5) $$
where $\Gamma$ is the common decay width for the two states,
$M_o$ is their mean mass and $\Delta m $ is the mass difference.
The time evolution (QQ3.5) can be viewed as the precession of a Q-spin
of $1/2$ around a magnetic field in Q-spin space with a precession
frequency $\Delta m$.
This formulation has a very simple physical interpretation.
Strong interactions conserve both $CP$ and flavor.
In the limit where weak interactions are neglected the flavor eigenstates
$B^o$ and $\bar B^o$ are degenerate and $Q$-spin is a good symmetry.
The weak interactions break the $Q$-spin symmetry,
but in the approximation where the difference between the lifetimes of
the two mass eigenstates can be neglected, the
$Q$-spin symmetry remains for rotations about the $x$ axis.
The symmetry breaking appears in this formulation as
a magnetic field in quasispin space in the direction of the x axis.
The mass eigenstates are then just the states with
``spin up" and ``spin down" with respect to this axis.
An initial state $ \ket {B(0)}$ which is not a mass eigenstate has
its spin pointing in a direction at some angle $\theta$ from the $x$
axis. This state evolves in time by having its spin precess with a
frequency $\Delta m$ in a path described by a cone at an angle $\theta$
around the $x$ axis.
\smallskip
\noindent{\twelvepoint 3.2\ CP Violation Described by Angles in Quasispin
Space}

\noindent
Any given decay mode; e.g. $ \psi K_S$ defines an axis in quasispin
space. A basis $(B_\mu,B_\nu)$ can always be defined in which the decay
$B_\mu \rightarrow f$ is allowed and $B_\nu \rightarrow f$ is forbidden.
We can always choose the parameters $\alpha$ and $\theta$ in the
notation (3.8) to make the transition matrix element
$\bra{f}T\ket{B_\nu} $ vanish for any final state $\ket {f}$. If
$$
e^{i\theta} \tan({{\alpha}\over{2}}) =
{{\bra{f}T\ket{\bar B^o}}\over{\bra{f}T\ket{B^o}}}
\eqno (QQ3.6a) $$
$$
e^{i{{\theta}\over{2}}} \sin({{\alpha}\over{2}}) \bra{f}T \ket{B^o} -
e^{-i{{\theta}\over{2}}} \cos({{\alpha}\over{2}}) \bra{f}T\ket{\bar B^o}
= $$ $$ =
\bra{f}T\ket{B_\nu} = 0 \eqno (QQ3.6b) $$
Thus every decay mode chooses a direction in quasispin space.

\noindent
If CP is conserved and $\ket{f}$ is a CP eigenstate, $B_\mu$ and $B_\nu$
are both CP and mass
eigenstates and $\cos(\alpha/2) = \sin(\alpha/2) = 1/\sqrt 2$.
We choose phases such that $\theta = 0$,
$B_\nu = B_1$ and is odd under $C$ while
$ B_\mu = B_2$ and is even under $C$,
The decays $B_1 \rightarrow \pi \pi$ and $B_2 \rightarrow \psi K_S$
are allowed and $B_1 \rightarrow \psi K_S$ and $B_2 \rightarrow \pi \pi$
are forbidden. The matrix $q_x$ is just the charge conjugation matrix in
this $ 2 \times 2 $ subspace, since its eigenstates are the C eigenstates
and its eigenvalues are the C eigenvalues.
\noindent

When $CP$ is violated, the $CP$ eigenstates are no longer well defined,
and the states $(B_\mu,B_\nu)$ are not necessarily mass eigenstates.
Consider the case with no final state rescattering in
the decays to $ \psi K_S$ and the Golden Rule applies to these decays.
Then
$|\bra{\psi K_S}T\ket{\bar B^o}| = |\bra{\psi K_S}T\ket{B^o}|$ and
$\tan(\alpha/2) = 1$. Then
$$ \ket{B_\mu} = (1/\sqrt 2)(e^{i{{\theta}\over{2}}}  \ket{B^o} +
e^{-i{{\theta}\over{2}}}   \ket{\bar B^o})
= e^{iq_z{{\theta}\over{2}}}\ket{B_2}
\eqno (QQ3.7a) $$
$$ \ket{B_\nu} = (1/\sqrt 2)(e^{i{{\theta}\over{2}}}   \ket{B^o}
- e^{-i{{\theta}\over{2}}}  \ket{\bar B^o})
= e^{iq_z{{\theta}\over{2}}}\ket{B_1}
\eqno (QQ3.7b) $$
The operator $q_z/2$ is just the generator of rotations about the
$z$ axis in quasispin space. Thus
the states $(B_\mu,B_\nu)$ are seen to differ from the mass
eigenstates by a quasispin rotation by an angle $\theta$ about the $z$
axis, and
$$
\vbox{\eqalignno{
 q_z \ket{B_\mu} = & \ket{B_\nu};    ~ ~ ~ ~ ~
 q_z \ket{B_\nu} = \ket{B_\mu} ~ ~ ~ &(QQ3.8a) \cr
\bra {B_\mu} q_z \ket {B_\mu} & =  \bra {B_\nu} q_z \ket {B_\nu} = 0
& (QQ3.8b)  \cr
\bra {B_\nu} q_y \ket {B_\nu} & =  -\bra {B_\mu} q_y \ket {B_\mu} =
\sin \theta
& (QQ3.8c)  \cr}} $$

\noindent
The angle $\theta$ defines a measure of $CP$ violation.
We can define a basis
$(B_\mu,B_\nu)$ and an angle $\theta$ for each decay mode into a $CP$
eigenstate and ask whether the bases and the value of $\theta$ are
the same for two different $CP$ eigenstates, like $\pi^+ \pi^-$ and
$\pi^o \pi^o$. This interesting physical question is related in
the analogous kaon case to the question of whether the parameter
$\epsilon'$ is zero. If all the parameters analogous to $\epsilon'$ are
zero, as in the case of the superweak theory, then $CP$ can be defined
to be conserved in decays to $CP$ eigenstates and unique $CP$ eigenstates
can be defined with a unique value of $\theta$.
If on the other hand, the parameter
analogous to $\epsilon'$ is not zero, then the angles $\theta$ defined
for different decay modes like $ \pi^+ \pi^-$ and $ \pi^o \pi^o$
will be different and there will be no unique definition of $CP$
eigenstates in the $B^o - \bar B^o$ space.

\noindent
In this approximation where the lifetime difference is
neglected, the degree of CP violation defined by a given decay mode and
the mass eigenstates is expressed by a single parameter $\sin \theta$.
This differs from the
kaon system which has two independent mechanisms for
mixing; namely the mass difference and the width difference
between the two eigenstates, and both the real and imaginary parts of the
complex parameter $\epsilon$
are needed for a complete description of the mixing and CP violation.
The exact relation between our parameter $\sin \theta$ and the
complex parameter $\epsilon$ in the conventional
formulation\REF{\Bigi}{I. I. Bigi and A. I Sanda, Nuc. Phys. {\bf B281},
41 (1987).}$[{\Bigi}]$
is given in refs.$\({\Bspin,\PEPRspin}\)$.
In the kaon case, the two eigenstates $K_L$ and $K_S$
disappear from the system at different rates,
the time evolution operator is not unitary because probability is not
conserved, and the eigenstates are not necessarily orthogonal.
This does not occur when the lifetime difference is
neglected and the time evolution operator can be factorized into a common
exponential decay and a unitary $2\times 2$ time-dependent matrix with
orthogonal eigenstates.
\smallskip
\noindent{\twelvepoint 3.3\ Lepton Asymmetries and CP Violation}

\noindent
In many experiments a state
$\ket{B_\nu}$
is prepared in some way and
a lepton asymmetry is measured; i.e.
the relative probability of decays
$W (B_\nu \rightarrow \mu^{\pm} + X) $
into positive or negative leptons.
$$ A_{lept}(B_\nu) \equiv
{{W (B_\nu \rightarrow \mu^{+} X) - W (B_\nu \rightarrow \mu^{-} X)}\over
{W (B_\nu \rightarrow \mu^{+} X) + W (B_\nu \rightarrow \mu^{-} X)}}
\eqno (QQ3.9a) $$
This lepton asymmetry is given by
the difference between the probability that the state
$\ket{B_\nu}$ is in a $B^o$ state or a $\bar B^o$ state. This is
just the expectation value of $q_z$ in the state $\ket{B_\nu}$, or
the ``polarization" in the z-direction in q-spin space.
$$ \vbox{\eqalignno{
A_{lept}(B_\nu) = &
{{|\langle B^o \ket{B_\nu}|^2 - |\langle \bar B^o \ket{B_\nu}|^2}\over
{|\langle B^o \ket{B_\nu}|^2 + |\langle \bar B^o \ket{B_\nu}|^2}} =
\cr
= & \bra{B_\nu} q_z \ket{B_\nu}
& (QQ3.9b)
\cr}} $$
Combining eqs. (QQ3.5) and (QQ3.9b) immediately gives the result of a
lepton asymmetry experiment on a state $ \ket {B(t)} $ prepared by
creating a state $ \ket {B(0)}$ at time $t=0$ and observing the lepton
asymmetry at time $t$,
$$ \vbox{\eqalignno{
& A_{lept}\{B(t)\} =
\bra{B(t)} q_z \ket{B(t)} =&
\cr
=
\bra {B(0)}  e^{i{{\Delta m}\over{2}}\cdot q_x t} & q_z
e^{-i{{\Delta m}\over{2}}\cdot
q_x t} \ket {B(0)} = &
\cr
=
\bra {B(0)}   q_z \cos(\Delta &m t) + q_y \sin(\Delta m t)  \ket {B(0)}
& (QQ3.10)  \cr}} $$
The time evolution of the expectation value of the
operator $q_z$ around this axis is seen from eq. (QQ3.10) to be
also expressed by rotating the
operator $q_z$ and calculating its expectation value in the state
$ \ket {B(0)}$.

\noindent
We immediately obtain the result of the usual $B^o - \bar B^o$
oscillations measured by lepton asymmetry by substituting these
states for $ \ket {B(0)}$ in eq. (QQ3.10) and using (QQ3.1d)
$$ \vbox{\eqalignno{
A_{lept}\{B^o(t)\} = & \cos(\Delta m t)
& (QQ3.11a)  \cr
A_{lept}\{\bar B^o(t)\} = & - \cos(\Delta m t)
& (QQ3.11b)  \cr}} $$

\noindent
We can also calculate the lepton asymmetry observed at time $t$ when
any state described in the form (QQ3.10) is created at time $t=0$.
Eqs. (QQ3.10) and (QQ3.11) give
$$ \vbox{\eqalignno{
A_{lept}\{B_\mu(t)\} &= \bra {B_\mu} q_y \sin(\Delta m t)  \ket {B_\mu}
=& \cr
&= - \sin \theta \sin(\Delta m t)
& (QQ3.12a)  \cr
A_{lept}\{B_\nu(t)\} &= \bra {B_\nu} q_y \sin(\Delta m t)  \ket {B_\nu}
=& \cr
&= \sin \theta \sin(\Delta m t)
& (QQ3.12b)  \cr}} $$
where we have used the expectation values (QQ3.4b)

\noindent
One of the reasons for discussing asymmetric B factories arises from the
necessity to measure time intervals. The lepton asymmetries (QQ3.12a) are
seen to be an odd function of the time $t$. In an experiment which gives
equal weight to positive and negative values of $t$, the lepton
asymmetry vanishes. We shall see below that this occurs in experiments
where a $B^o - \bar B^o$
pair is produced from the decay of an $\Upsilon(4S)$ and one is
observed to decay into a $CP$ eigenstate and the other into leptons.
The results (QQ3.12) will be shown below to apply to such an experiment
not only for the case where the leptonic decay decays occur after the
decay into the $CP$ eigenstate and $t \geq 0$ but also
when the leptonic decay $precedes$ the decay into the $CP$ eigenstate.
In the latter case the results (QQ3.12) apply with $t \leq 0$
Thus in any experiment which gives equal weight to these two
cases and does not measure the sign of the time interval the
$CP$-violation asymmetry is lost.

\noindent
We can obtain a general view of how CP violation is observed
experimentally from the picture of how a unit vector $\hat z$
originally in the positive $z$ direction rotates in time about the
$x$ axis. Let us choose
the sign of $\Delta m$ so that the direction of rotation is
$$ + \hat z \rightarrow - \hat y \rightarrow - \hat z \rightarrow
+ \hat y \rightarrow + \hat z \eqno(QQ3.13a)     $$
It is convenient to define a basis
(QQ3.7) with the angle $\theta$ chosen so that
$B_{\mu}$ decays into a chosen $CP$ eigenstate; e.g. $\psi K_S$, and the
amplitude vanishes for the $B_{\nu}$ decay into this mode.
Let us now consider the case $\sin \theta = 1$ which gives maximum $CP$
violation. Then the states $B_{\mu}$ and $B_{\nu}$ are seen from
eq. (QQ3.8c) to be eigenvectors of $q_y$ with the eigenvalues $\mp 1$.
The state $B^o$ which points in the direction $+z$ will rotate
$$ B^o
\rightarrow B_\mu (\rightarrow \psi K_S)
\rightarrow \bar B^o
\rightarrow B_\nu (\rm no ~ \psi K_S)
\rightarrow B^o
\eqno(QQ3.13b)     $$
We immediately see that for values of $t$ in the first half period
where
$\Delta mt \leq \pi$ and $\sin (\Delta mt) \geq 0$
a state which is originally $B_\mu$ rotates in the direction toward
$\bar B^o$ and gives a negative lepton asymmetry, while
a state which is originally $B_\nu$ rotates in the direction of
$B^o$ and gives a positive lepton asymmetry, in agreement with
the results (QQ3.12).

\noindent
In any experiment where
$B$ pairs are produced in a $C$ eigenstate with a negative
eigenvalue, as in the decay of the $\Upsilon(4S)$, the oscillations of
the pair are correlated until the time of the first decay.
Angular momentum is conserved during the oscillation; thus a pair
in an odd-C eigenstate remains until the first decay
in a state of odd angular momentum (a
$p$-wave in the case of the $\Upsilon$ decays) which is antisymmetric in
space and forbidden for identical bosons. Thus
if the first decay is into $\psi K_S$, the second must be in the state
$B_\nu$ and a positive lepton asymmetry will be observed during the
first half period after the first decay.
However, if the first decay is in a leptonic mode we can see that
a negative lepton asymmetry will be observed during the first half
period after the first decay.
If the leptonic mode defines the decaying
state as $B^o$, the other is $\bar B^o$ and rotates during the first
half period into the state $B_\nu$ which does not decay into $\psi K_S$.
If the leptonic mode defines the state as
state as $\bar B^o$, the other is $B^o$ and rotates during the first
half period into the state $B_\mu$ which does decay into $\psi K_S$.
Thus the second decay chooses states emitted originally with $\bar B^o$.

\noindent
Thus interchanging the times of the two
decays reverses the $CP$ asymmetry as indicated by using both positive
and negative times in the expressions (QQ3.12).
If no time measurement is made in the experiment, the events where the
leptonic decays occur before and after the $\psi K_S$ decay are added
and the $CP$-violating asymmetry cancels and is not observed.

\noindent
Time measurements are not feasible when the $\Upsilon(4S)$
is produced at rest, because the $B$ mesons have very low momentum and
the path between production and decay is too small to be measured by
vertex detectors. Therefore measurements of this kind of asymmetry
require an ``asymmetric $B$ factory" in which the $\Upsilon(4S)$
is produced in flight and time measurements can be made with vertex
detectors.

\noindent
This time symmetry effect can be seen quantitatively
by some elementary quasispin
properties of the amplitude $\langle B_\mu \ket{B^o(t)}$ describing the
probability that a B meson in the state $B^o$ after a time $t$ is in the
state $B_\mu$ and be observed to decay into a given $CP$ eigenstate. This
describes an experiment in which a $B^o - \bar B^o$ pair is produced,
the $\bar B^o$ decays at time $t=0$ in a leptonic mode identifying it as
$\bar B^o$ and therefore requiring the other meson to be a $B^o$ at
$t=0$, and the second meson decays at time $t$.
$$ \vbox{\eqalignno{
& e^{{\Gamma\over 2}t}\cdot   \langle B_\mu \ket{B^o(t)}
= \bra{B_\mu} e^{-i{{\Delta m}\over{2}}\cdot q_x t} q_z \ket{B^o}
&= \cr
&= \bra{B_\mu} q_z e^{i{{\Delta m}\over{2}}\cdot q_x t}  \ket{B^o}
= \bra{B_\nu} e^{i{{\Delta m}\over{2}}\cdot q_x t}  \ket{B^o}
&= \cr
&= \bra{B^o} e^{i{{\Delta m}\over{2}}\cdot q_x t}  \ket{B_\nu}^*
& (QQ3.14a)  \cr
&- e^{{\Gamma\over 2} t}\cdot  \langle B_\mu \ket{\bar B^o(t)}
= \bra{B_\mu} e^{-i{{\Delta m}\over{2}}\cdot q_x t} q_z \ket{\bar B^o }
&= \cr
&= \bra{B_\mu} q_z e^{i{{\Delta m}\over{2}}\cdot q_x t}  \ket{\bar B^o}
= \bra{B_\nu} e^{i{{\Delta m}\over{2}}\cdot q_x t}  \ket{\bar B^o}
&= \cr
&= \bra{\bar B^o} e^{-i{{\Delta m}\over{2}}\cdot q_x t}  \ket{B_\nu}^*
& (QQ3.14b)  \cr}} $$
where we have used eqs. (QQ3.1a) and (QQ3.8a) and noted that the
quasispin operator $q_z$ anticommutes with $q_x$.
Thus the probability that a state
created as $\ket{B^o}$ at time t=0 is observed as the state
$\ket{B_\mu}$ at time $t$ is equal to the probability that a state
created as $\ket{B_\nu}$ at time $t=0$  is observed as the state
$\ket{B^o}$ at time $t$. The latter can
describe an experiment in which a $B^o - \bar B^o$ pair is produced
and is an antisymmetric state like that produced in the decay of the
$\Upsilon(4S)$ at time $t=0$.
One meson decays at time $t=0$ into a $CP$ eigenstate like $\psi K_S$
and therefore requires the other meson to be a $B_\nu$ at
$t=0$, and the second meson decays at time $t$ leptonically in a mode
allowed for $B^o$.

\noindent
In both cases one meson decays into a $CP$ eigenstate, and the
probability that the other decays like a $B^o$ at a time $t$ $after$ the
decay into the $CP$ eigenstate is seen to be equal to the probability
that the other decays like a $\bar B^o$ at a time $t$ $before$ the
decay into the $CP$ eigenstate.
Thus we again see that interchanging the times of the two
decays reverses the $CP$ asymmetry.
\smallskip
\noindent{\twelvepoint 3.4\ Measuring $CP$ Violation by Observing the
Decay of a Tagged neutral $B$ meson into a $CP$ eigenstate}

\noindent
A typical $CP$-violation experiment using neutral $B$ mesons involves
measuring the angle $\theta$ defined by eqs. (QQ3.2) between the axis in
quasispin space defined by a particular decay mode and the $x$ axis
defined by the mass eigenstates. We again use the basis
(QQ3.7) with the angle $\theta$ chosen so that
$B_{\mu}$ decays into the measured decay mode; e.g. $\psi K_S$ and the
amplitude vanishes for the $B_{\nu}$ decay into this mode.
In these experiments the decay of
a neutral $B$ is observed some time after it has been
``tagged" by observing another particle created together with it
which identifies it as a definite state in the basis (QQ3.7).
The neutral $B$ oscillates
in time according to eqs. (QQ3.5) and (QQ8.1b) during the interval
between the tagging and decay times.

\smallskip
\noindent{\twelvepoint 3.4.1 \ {\it Tagging by a Charged B}}

\noindent
The simplest tagging method uses a charged $B \bar B$ pair, since the
charged $B$ does not oscillate and the observation of a $B^+$ or $B^-$ at
time $t=0$ automatically defines its partner as $\bar B^o$ or $B^o$ at
t=0. If the decay is observed at time $t$ into $ \psi K_S$ the decay
amplitude depends on the overlap of the oscillating state at time $t$
with the state $B_\mu$ (QQ3.2a) defined by the $ \psi K_S$ decay The
$CP$ asymmetry is given by the difference between the squares of these
overlaps for initial states tagged as $\bar B^o$ or $B^o$ at t=0.
$$ \vbox{\eqalignno{
& A_{CP}(t) \equiv {{N(t) - \bar N(t)}\over{N(t) + \bar N(t)}} =
& \cr
& = {{
|\langle B_\mu \ket{B^o(t)}|^2 - |\langle B_\mu \ket{\bar B^o(t)}|^2
}\over{
|\langle B_\mu \ket{B^o(t)}|^2 + |\langle B_\mu \ket{\bar B^o(t)}|^2
}}
& (QQ3.15a)  \cr}} $$
where $N(t)$ and $\bar N(t)$ denote the decay rates into the measured
$CP$ eigenstates for initial states tagged respectively as
as $B^o$ or $\bar B^o$ at t=0. Then
$$ \vbox{\eqalignno{
& A_{CP}(t)
= & \cr
& = |\bra{B_\mu}
e^{-i{{\Delta m}\over{2}}\cdot q_x t} \ket{B}|^2 - |\bra{B_\mu}
e^{-i{{\Delta m}\over{2}}\cdot q_x t} \ket{\bar B}|^2
& \cr
& =
\bra{B_\mu} e^{-i{{\Delta m}\over{2}}\cdot q_x t} q_z
e^{i{{\Delta m}\over{2}}\cdot q_x t} \ket{B_\mu}
& \cr
& = - \bra{B_\mu} q_y \ket{B_\mu} \sin (\Delta m t)
& \cr
& = \sin \theta \sin(\Delta m t)
& (QQ3.15b)  \cr}} $$
Then
$$ {{
\int_0^\infty N(t)dt - \int_0^\infty \bar N(t)dt
}\over{
\int_0^\infty N(t)dt + \int_0^\infty \bar N(t)dt
}} = \sin \theta \cdot
{{\Delta m \Gamma}\over{\Gamma^2 + (\Delta m)^2}}
\eqno(QQ3.16)      $$
Thus the $CP$ violation asymmetry remains when there is no time
measurement and the observed data give the time integrals (3.17).

\smallskip
\noindent{\twelvepoint 3.4.2 \ {\it Tagging by Coherent Decays of
Neutral B's Produced by $\Upsilon(4S)$ decays}}

\noindent
Tagging of neutral $B$ pairs is more complicated than tagging by a
charged $B$ because both neutral $B$ mesons oscillate with time, and CP
violation asymmetries will depend in general on two time intervals.
The general approach is to measure one decay into a mode like a leptonic
mode, which is allowed only for $B^o$ or $\bar B^o$ at a time which we
denote by $t_{\pm}$ and the other into
a $CP$ eigenstate, which identifies the decaying state as $B_\mu$, at
a time which we denote by $t_\mu$. We define the $CP$ asymmetry as
$$ A_{CP}(t_\pm,t_\mu) \equiv
{{
N(t_\pm,t_\mu) - \bar N(t_\pm,t_\mu)
}\over{N(t_\pm,t_\mu) + \bar N(t_\pm,t_\mu)
}}
\eqno(QQ3.17)      $$
where $N(t_\pm,t_\mu)$ and $\bar N(t_\pm,t_\mu)$
denote the decay rates into the measured $CP$ eigenstates at time $t_\mu$
together with a decay allowed respectively for $B^o$ or $\bar B^o$
at time $t_\pm$.

\noindent
When the first decay is observed in a mode allowed only for $B^o$ or
$\bar B^o$, $t_\pm \leq t_\mu$ and
the second $B$ must be in the state $\bar B^o$ at time $t_\pm$
if the observed first decay is into a mode allowed for $B^o$,
and vice versa. The second $B$ thus behaves
in the same way as if it were tagged by a charged $B$ as above.
We can immediately use the result (QQ3.15b) with a sign reversal since
here $N$ refers to the case where the second $B$ is tagged as a
$\bar B^o$ at time $t_\pm$ and
$\bar N$ refers to the case where the second $B$ is tagged as a $B^o$.
\noindent
$$ A_{CP} (t_\pm \leq t_\mu)
= - \sin \theta \sin\{\Delta m (t_\mu - t_\pm)\}
\eqno(QQ3.18a)      $$
When the first decay is observed in the $CP$ eigenstate allowed for
the state $B_\mu$, $t_\pm \geq t_\mu$ and
the other neutral $B$ is immediately identified as
$B_\nu$ at time $t_\mu$. The $CP$ asymmetry for this case is immediately
given by eq. (QQ3.12b)
$$ \vbox{\eqalignno{
A_{CP}(t_\pm \geq t_\mu) & =
\sin \theta \sin\{\Delta m (t_\pm - t_\mu)\} =&
\cr
= - \sin \theta & \sin\{\Delta m (t_\mu - t_\pm)\}
& (QQ3.18b)
\cr}} $$
We again see that the the same odd function
$ - \sin \theta \sin\{\Delta m (t_\mu - t_\pm)\} $
holds for both positive and negative values of $t_\mu - t_\pm $.
If the decay of
two neutral $B$ mesons produced from the decay of an $\Upsilon(4S)$ are
observed, and there is no time measurement to determine which decay
occurred first, the observed asymmetry will be
given by the sum of the results (QQ3.18a) and (QQ3.18b) and will exactly
cancel.
\medskip
\noindent{\twelvepoint 4.\ {\caps The Difference
Between $B$ and $K$ Physics - The Good News and the Bad News}}

\noindent
We now return to our two initial questions and summarize
expectations from $B$ physics and $B$ factories.
\smallskip
\noindent{\twelvepoint 4.1\ No Dominant $B$ Decay Mode}

\noindent
Kaon decay is dominated by the $2\pi$ final state which is a $CP$
eigenstate and has much larger phase space than any other decay mode.
There is no such dominant mode in $B$ decays. Nearly all decays go to
final states having nontrivial naked charm or strangeness and therefore
occur in pairs, one allowed for $B$ and the other for $\bar B$.
\smallskip
\noindent{\twelvepoint 4.1.1 \ {\it No Lifetime Difference}}

\noindent
The $K_L - K_S$ lifetime difference arises from the dominant decay $K_S
\rightarrow 2\pi$ which determines the $K_S$ lifetime and has no
counterpart in $K_L$. In $B$ decays the dominant modes occur in pairs
which will contribute equally to the decay rates of the two mass
eigenstates. Thus the $B$ mass eigenstates have a negligible lifetime
difference.
\smallskip
\noindent{\twelvepoint 4.1.2 \ {\it
 Mass Eigenstates Not Separated by Waiting}}

\noindent
The lifetme difference in the kaon case allows separating the two mass
eigenstates simply by waiting. An essentially pure $K_L$ beam is
obtained by waiting a sufficient number of $K_S$ lifetimes, and $CP$
violation is then observed by decays of $K_L$ into a $CP$ eigenstate
with the wrong eigenvalue. This is impossible in $B$ decays, where  the
two mass eigenstates have equal lifetimes.
\smallskip
\noindent{\twelvepoint 4.2\ Many $B$ Decay Modes}
\smallskip
\noindent{\twelvepoint 4.2.1 \ {\it
 Rich Data - Small Branching Ratios $\approx$ 1\%}}

\noindent
There are many $B$ decay modes that can be studied and can give different
information about weak interactions and $CP$ violation. However, each
individual mode has a small branching ratio, and $CP$ eigenstates like
$\psi K_S$ which are of particular interest have small branching ratios.
\smallskip
\noindent{\twelvepoint 4.2.2 \ {\it
 Final State Rescattering - Beats Golden Rule}}

\noindent
The large number of final states in $B$ decays provide opportunities for
strong rescattering processes and thereby beating the golden rule
restriction on charge asymmetries when there is no rescattering.
\smallskip
\noindent{\twelvepoint 4.2.3 \ {\it
 Conspiracies Beat CPT Restrictions}}

\noindent
The large number of decay modes allow for conspiracies in which opposite
$CP$-violating asymmetries are observed for different modes and the sum
of the asymmetries in all decay modes adds up to zero to give the same
total width for both $B^o$ and $\bar B^o$.
\smallskip
\noindent{\twelvepoint 4.3\
 $B^o - \bar B^o$ Oscillations During Decay}

\noindent
The equal lifetimes enables oscillations to be observed.
\smallskip
\noindent{\twelvepoint 4.3.1 \ {\it
Time Dependence Confuses Measurements}}

\noindent
When a $B^o - \bar B^o$ pair is produced coherently, both oscillate in
time and tagging one $B$ by observing the decay of the other is simple
only when time measurements are made. Such time measurements cannot be
made on $B$'s produced in decays of the $\Upsilon (4S)$ at rest
because the mass of the $\Upsilon(4S)$ is just above the $B \bar B$
threshold.
\smallskip
\noindent{\twelvepoint 4.3.2 \ {\it
CP Violation Observable in Mixing Phases }}

\noindent
The $B^o - \bar B^o$ mixing has a complex phase which is related to $CP$
violation. In the quasispin notation this mixing chooses the $x$
direction in quasispin space. If $CP$ is violated this direction can be
different from the direction chosen by the decay into a $CP$ eigenstate
like $\psi K_S$ and the angle between the two directions measured in an
experiment.
\smallskip
\noindent{\twelvepoint 4.4\
 All Dominant Hadronic B decays involve 3 Generations}

\noindent
In the standard model with the CKM matrix, three generations are needed
to observe $CP$ violation. Thus all processes which can be described in
terms of diagrams involving only 2 generations cannot show $CP$
violation.
\smallskip
\noindent{\twelvepoint 4.4.1 \ {\it
CP Violation Observable in charm and strangeness decays only via
diagrams with virtual t and b quarks}}

\noindent
Charm and strange decays which are described completely by diagrams
containing only $u$, $d$, $s$ and $c$ quarks do not show $CP$ violation
in the standard model. In these cases $CP$ violation is observable only
via contributions from diagrams containing virtual $t$ and $b$ quarks;
e.g. in the box diagram responsible for $K^o - \bar K^o$ mixing.
\smallskip
\noindent{\twelvepoint 4.4.2 \ {\it
 CP violation Observable in B Decays in Direct Diagrams}}

\noindent
In $B$ decays three generations are nearly always present in the direct
decay diagram. The favored vertex at the quark level
$b \rightarrow c d \bar u$ already contains flavors belonging to three
generations. Thus there are more possibilities of observing $CP$
violation in $B$ decays than in $K$ or $D$ decays.
\medskip
\noindent{\twelvepoint 5. \  {\caps Conclusion - The Lipkin Approach to
CP}}

\noindent
So far there has been no experimental evidence for $CP$ violation
outside of kaon physics. Any
\noindent
indication of CP violation in $B$ physics
would be a great breakthrough. It will be a long time before we have
enough good clear data to test the standard model predictions for $CP$
violation in $B$ decay. Inadequate data are always available before
adequate data. Our goal must be to get the maximum information from the
available data at each stage and to use this information to plan
subsequent experiments. Keep the standard model in mind but try to use a
more general approach. Look for easy experiments that even Lipkin can do
- even if theorists say no.

\noindent
There are many questions that can be investigated with early data and
which can be useful for future plans. Some examples are:

1. Is there $CP$ violation in $B$ physics?

2. What is the ball park of $CP$ violation?

3. What are the branching ratios for
physically interesting final states like $CP$ eigenstates?

4. Are there additional $CP$ eigenstates not yet observed that can
be useful?
\spointbegin States containing $\eta_c$ and other charmonium states.
\spoint States like $\psi K^*$ where different partial waves have
different $CP$ eigensvalues - perhaps one partial wave is dominant.
\spoint States like $K_S \psi' \rightarrow K_S \pi^+ \pi^- X $ where
the particle $X$ is not observed but can be identified by missing mass
kinematics.

5. How can one estimate penguin diagram contributions?

\refout

\end